\documentclass[twocolumn,aps]{revtex4}

\usepackage{graphicx}

\newcommand{\be}{\begin{equation}}
\newcommand{\ee}{\end{equation}}
\newcommand{\ben}{\begin{eqnarray}}
\newcommand{\een}{\end{eqnarray}}

\begin{document}
\title{Confinement in the 3-dimensional Gross-Neveu model}
\author{A. P. C. Malbouisson}
\address{{\it Centro Brasileiro de Pesquisas F\'{\i}sicas,
Rua Dr. X. Sigaud 150, 22290-180, Rio de Janeiro, RJ, Brazil}}
\author{J. M. C. Malbouisson, A. E. Santana}
\address{{\it Instituto de F\'{\i}sica, Universidade Federal da
Bahia, 40210-340, Salvador, BA, Brazil}}
\author{J. C. da Silva}
\address{{\it Centro Federal de Educa\c c\~ao Tecnol\'ogica da Bahia,
40300-010, Salvador, BA, Brazil}}

\begin{abstract}
We consider the $N$-components $3$-dimensional massive Gross-Neveu
model compactified in one spatial direction, the system being
constrained to a slab of thickness $L$. We derive a closed formula
for the effective renormalized $L$-dependent coupling constant in
the large-$N$ limit, using bag-model boundary conditions. For
values of the fixed coupling constant in absence of boundaries
$\lambda \geq \lambda_c \simeq 19.16$, we obtain ultra-violet
asymptotic freedom (for $L \rightarrow 0$) and confinement for a
length $L^{(c)}$ such that $2.07\, m^{-1} < L^{(c)} \lesssim
2.82\, m^{-1}$, $m$ being the fermionic mass. Taking for $m$ an
average of the masses of the quarks composing the proton, we
obtain a confining legth $L^{(c)}_p$ which is comparable with an
estimated proton diameter.
\newline
\newline
\noindent PACS number(s): 11.10.Jj; 11.10.Kk; 11.15.Pg
\end{abstract}

\maketitle

The currently accepted theory of strong interactions, Quantum
Chromodynamics (QCD), has an intricate structure which makes it
difficult to be directly handled. For this reason, simplified
models have been largely employed over the last decades in trials
to obtain clues about the behavior of interacting quarks. Among
them one of the most celebrated is the Gross-Neveu model
\cite{GN}, which considers the direct four-fermions interaction.
From a physical viewpoint, any satisfactory model describing
strong interactions must account for the known phenomenon of
confinement of the elementary constituents ({\it quarks}) of
hadrons. Such a property has not yet been analytically proved,
although lattice QCD gives clear indications that it indeed
happens.

In this Letter, we consider the $N$-component tridimensional
massive Gross-Neveu model, in the large-$N$ limit, compactified
along one of the spatial directions. For weak coupling the model
presents only asymptotic freedom, as expected. For strong
coupling, we analytically demonstrate the simultaneous existence
of ultra-violet (short distance) asymptotic freedom and
confinement. The confining length obtained is comparable with an
estimated classical proton diameter.

Although perturbatively non renormalizable for dimensions $D>2$,
the massive Gross-Neveu model in tridimensional space has been
shown to exist and has been constructed in Ref.~\cite{Jarrao}. The
idea that perturbative renormalizability should not be an absolute
criterion to a model to be physically consistent comes from works
done more than one decade ago. The results in renormalization
theory reported in Refs. \cite{Gaw1,Gaw2} give a theoretical
foundation to support the idea that non perturbatively
renormalizable models can exist and have a physical meaning
\cite{Weinberg1}. In fact, the Gross-Neveu model for dimensions
greater than $D=2$ has been investigated since the middle of the
seventies. Since then we know that the difficulties related to the
perturbative non-renormalizability of the model can be surmounted,
at least for $D=3$, by performing partial summations in the bare
perturbation series, while maintaining its initial structure (see
Refs. \cite {GN,parisi,anp}). This point has been also studied in
Ref.~\cite{rwp1}. The essence of the method lies in the fact that
the model in three dimensions may acquire a naive power counting
behavior of a just renormalizable theory, if we perform a partial
summation, to all orders in the coupling constant, of the chains
of one-loop bubbles before trying to renormalize the perturbation
series directly.

In present work, we perform such a summation for the large-$N$
model and we renormalize it employing a modified minimal
subtraction scheme, where the coupling constant counter-term is a
pole at the physical value of the argument of generalized Epstein
$zeta$-functions \cite{Epstein}. This technique, which employs
analytic and $zeta$-function regularizations, has been recently
used to study the boundary dependence of the coupling constant and
the spontaneous symmetry breaking in the compactified $\phi^4$
model \cite{JMario,Ademir}.

We start from the Wick-ordered massive Gross-Neveu Lagrangian in
Euclidian metrics (in relativistic units, $\hbar = c = 1$),
\begin{equation}
{\cal L}=:\bar{\psi}(x)(i\gamma^{\mu}\partial_{\mu}+m)\psi
(x):+\frac{u}{2}:\left[\bar{\psi} (x)\psi (x)\right]^{2}:,
\label{GN}
\end{equation}
where $m$ is the mass, $u$ the coupling constant, $x$ is a point
of ${\bf R} ^{D}$ and the $\gamma $-matrices are either $2\times
2$ or $4\times 4$. The fermionic field of spin $\frac{1}{2}$ has
$N$ components, $\psi ^{a}(x)$ $a=1,2,...,N$, and summation over
flavor and spin indices are understood. In the language of
Ref.~\cite{Jarrao} a theorem has been proved stating that a family
of parameters can be exhibited such that, for $N$ sufficiently
large, all renormalized Schwinger functions of the type
$S_{2p}(m,\lambda ,N)$, p=1,2,... exist. Thus we examine in the
following the large-$N$ zero external-momenta four-point function,
which is the basic object for our definition of the renormalized
coupling constant. The four-point function at leading order in
$\frac{1}{N}$ is given by the sum of all chains of single one-loop
diagrams. This sum gives for the four-point function at zero
external momenta the formal expression \cite{Jarrao}
\begin{equation}
\Gamma _{D}^{(4)}(0)=\;\frac{u}{1+Nu\Sigma (D)},  \label{4-point1}
\end{equation}
where $\Sigma (D)$ is the Feynman integral corresponding to the single
one-loop subdiagram
\begin{equation}
\Sigma (D)=\int \frac{d^{D}k}{(2\pi )^{D}}\;\frac{m^{2}-k^{2}}{\left(
k^{2}+m^{2}\right) ^{2}}.  \label{sigma02}
\end{equation}
Notice that $\Sigma (D) $ has dimension of $mass^{D-2}$ and the
coupling constant $u$ has dimension of $mass^{D-4}$.

We consider the system bounded between two parallel planes, normal
to the $x$ -axis, a distance $L$ apart from one another. We use
Cartesian coordinates $ {\bf r}=(x,{\bf z})$, where ${\bf z}$ is a
$(D-1)$-dimensional vector, with corresponding momenta ${\bf
k}=(k_{x},{\bf q})$, ${\bf q}$ being a $(D-1)$-dimensional vector
in momenta space. In addition, we assume that the field satisfies
bag-model boundary conditions \cite{Chodos,Lutken} on the planes
$x=0$ and $x=L$. These boundary conditions, constraining the
$x$-dependence of the field to a segment of length $L$, allow us
to proceed with respect to the $x$ -coordinate in a manner
analogous as it is done in the imaginary-time formalism in field
theory. This extension of the Matsubara procedure to describe a
model compactified in a space-coordinate has been already used in
the case of bosonic fields \cite{JMario,Ademir}. Accordingly, for
fermionic fields, the Feynman rules should be modified following
the prescription
\begin{equation}
\int \frac{dk_{x}}{2\pi }\rightarrow \frac{1}{L}\sum_{n=-\infty }^{+\infty
}\;,\;\;\;\;\;\;k_{x}\rightarrow \frac{2(n+\frac{1}{2})\pi }{L}\equiv \omega
_{n}\,.  \label{Matsubara}
\end{equation}
Then, the $L$-dependent four-point function at zero external momenta has the
formal expression,
\begin{equation}
\Gamma _{D}^{(4)}(0,L)=\;\frac{u}{1+Nu\Sigma (D,L)},  \label{4-point1L}
\end{equation}
where $\Sigma (D,L)$ is the $L$-dependent Feynman sum-integral
corresponding to the single one-loop subdiagram,
\begin{equation}
\Sigma (D,L)=\frac{1}{L}\sum_{n=-\infty }^{\infty }\int
\frac{d^{D-1}k}{ (2\pi )^{D-1}}\frac{m^{2}-({\bf k}^{2}+\omega
_{n}^{2})}{({\bf k}^{2}+\omega _{n}^{2}+m^{2})^{2}}.
\label{sigma0}
\end{equation}

Using the dimensionless quantities $q_{j}=k_{j}/2\pi m$ \thinspace
($ j=1,2,...,D-1$) and $a=(mL)^{-2}$, the single bubble $\Sigma
(D,L)$ can be written as
\begin{eqnarray}
\Sigma (D,a) &=&\left. \Sigma (D,a,s)\right| _{s=2}  \nonumber \\
&=& m^{D-2(s-1)}\sqrt{a} \nonumber \\
 && \times \left. \left[ \frac{1}{4\pi ^{2}}
U_{1}(s,a)-U_{2}(s,a)-aU_{3}(s,a)\right] \right| _{s=2},\nonumber \\
\label{sigmaA}
\end{eqnarray}
where
\begin{eqnarray}
U_{1}(s,a) &=&\frac{\pi ^{(D-1)/2-2s+2}}{2^{2s-2}}\frac{\Gamma
(s-(D-1)/2)}{
\Gamma (s)}  \nonumber \\
&&\times \sum_{n=-\infty }^{+\infty }\left[
a(n+\frac{1}{2})^{2}+(2\pi )^{-2} \right] ^{(D-1)/2-s} \nonumber \\
\label{UU1}
\end{eqnarray}
and
\begin{eqnarray}
U_{2}(s,a) &=&\frac{D-1}{2(s-1)}U_{1}(s-1,a),  \label{UU2} \\
U_{3}(s,a) &=&\frac{1}{1-s}\frac{\partial }{\partial a}U_{1}(s-1,a).
\label{UU3}
\end{eqnarray}
These expressions were obtained by performing the integrals over
$q_{j}$ in Eq.~(\ref{sigma0}) using well-known dimensional
regularization formulas \cite {Ramond}.

Sums of the type appearing in Eq.~(\ref{UU1}) can be casted in the
form
\begin{equation}
\sum_{n=-\infty }^{+\infty }\left[ a(n+\frac{1}{2})^{2}+c^{2}\right] ^{-\nu
}=4^{\nu }Z_{1}^{4c^{2}}(\nu ,a)-Z_{1}^{c^{2}}(\nu ,a),  \label{EF}
\end{equation}
where
\begin{equation}
Z_{1}^{b^{2}}(\nu ,a)=\sum_{n=-\infty }^{+\infty }\left[
an^{2}+b^{2}\right] ^{-\nu } \label{Es1}
\end{equation}
is one of the generalized Epstein $zeta$-functions \cite{Epstein}
defined for $\textrm{Re}\nu >1/2$, which possesses the following
analytic extension
\begin{eqnarray}
Z_{1}^{b^{2}}(\nu ,a) &=&\frac{\sqrt{\pi }}{\sqrt{a}\,\Gamma (\nu )}\left[
\frac{\Gamma (\nu -1/2)}{b^{2\nu -1}}\right.   \nonumber \\
&&\left. +4\sum_{n=1}^{\infty }\left( \frac{b\sqrt{a}}{\pi
n}\right) ^{\frac{ 1}{2}-\nu }K_{\frac{1}{2}-\nu }\left(
\frac{2\pi b n}{\sqrt{a}}\right) \right] , \nonumber \\
\label{extanali}
\end{eqnarray}
valid for the whole complex $\nu $-plane. Therefore, the functions
$ U_{l}(s,a)$ ($l=1,2,3$) can be also extended to the whole
complex $s$-plane, and we obtain
\begin{eqnarray}
U_{1}(s,a) &=&\frac{h(s,D)}{\sqrt{a}}\left[ \Gamma
(s-\frac{D}{2})+4W(s,a)
\right] ,  \label{UA1} \\
U_{2}(s,a) &=&\frac{(D-1)h(s-1,D)}{2(s-1)\sqrt{a}}  \nonumber \\
&&\times \left[ \Gamma (s-1-\frac{D}{2})+4W(s-1,a)\right] ,  \label{UA2} \\
U_{3}(s,a) &=&\frac{h(s-1,D)}{2(s-1)a\sqrt{a}}\left[ \Gamma
(s-1-\frac{D}{2}
)\right.   \nonumber \\
&&\left. +4W(s-1,a)-8a\frac{\partial }{\partial a}W(s-1,a)\right]
, \nonumber \\
\label{UA3}
\end{eqnarray}
where
\begin{equation}
h(s,D)=\frac{\pi ^{2-D/2}}{2^{D-2}\Gamma (s)}  \label{aga}
\end{equation}
and
\begin{eqnarray}
W(s,a) &=&2\sum_{n=1}^{\infty }\left( \frac{\sqrt{a}}{n}\right)
^{\frac{D}{2}
-s}K_{\frac{D}{2}-s}\left( \frac{2n}{\sqrt{a}}\right)   \nonumber \\
&&-\sum_{n=1}^{\infty }\left( \frac{2\sqrt{a}}{n}\right)
^{\frac{D}{2}-s}K_{ \frac{D}{2}-s}\left( \frac{n}{\sqrt{a}}\right)
.  \label{W}
\end{eqnarray}

Replacing these expressions in Eq.(\ref{sigmaA}), we see that
$\Sigma (D,L,s)$ can be written in the form
\begin{equation}
\Sigma (D,L,s)=H(D.s)+G(D,L,s),
\end{equation}
where $H(D,s)$ is a term independent of $a$, coming from the first
parcels between the brackets in Eqs.(\ref{UA1}), (\ref{UA2})and
(\ref{UA3}), while $ G(D,L,s)$ is the term arising from the sum
over the parcels containing the $W $-functions in these equations.
We are using a modified minimal subtraction scheme, where the
coupling constant counter-terms are poles appearing at the
physical value $s=2$. Then the $L$ -dependent correction to the
coupling constant is proportional to the regular part of the
analytical extension of the Epstein $zeta$ -function in the
neighborhood of the pole at $s=2$. We see from Eqs. (\ref {UA2})
and (\ref{UA3}) that, for $s=2$ and even dimensions $D\geq 2$,
$H(D)$ is divergent due to the pole of the $\Gamma $-function.
Accordingly this term should be subtracted to give the
renormalized single bubble function $ \Sigma _{R}(D,L)$. For sake
of uniformity, the term $H(D)$ is also subtracted in the case of
other dimensions $D$, where no poles of $\Gamma $-functions are
present; in these cases, we perform a finite renormalization.
Then, we simply get, for any dimension $D$,
\begin{eqnarray}
\Sigma _{R}(D,L) &=&m^{D-2(s-1)}\left[ \frac{h(s,D)}{\pi ^{2}}W(s,a)\right.
\nonumber \\
&&-\left. \frac{2Dh(s-1,D)}{s-1}W(s-1,a)\right.   \nonumber \\
&&+\left. \left. \frac{4h(s-1,D)a}{s-1}\frac{\partial }{\partial
a}W(s-1,a) \right] \right| _{s=2} . \nonumber \\ \label{sigmaR}
\end{eqnarray}

In dimension $D=3$, using the explicit form for the Bessel
functions $K_{\pm 1/2}(z)=\sqrt{\pi }e^{-z}/\sqrt{2z}$ and
performing the resulting sums, we obtain (remembering that
$a=(mL)^{-2}$)
\begin{eqnarray}
\Sigma _{R}(3,L) &=&m\left[ \frac{4\pi ^{2}+1}{2\pi }\left(
\frac{e^{-2mL}}{
1-e^{-2mL}}\right) \right.   \nonumber \\
&&-\left. \frac{4\pi ^{2}+1}{4\pi }\left( \frac{e^{-mL}}{1-e^{-mL}}\right)
\right.   \nonumber \\
&&+\left. \frac{2\pi }{mL}\log \left(
\frac{1-e^{-2mL}}{1-e^{-mL}}\right) \right] .  \label{sigmaR3}
\end{eqnarray}
Substituting this expression into Eq. (\ref{4-point1L}), we obtain
the large-$N$ effective renormalized coupling constant, taking
$Nu=\lambda $ fixed as usual,
\begin{equation}
g(3,L)=N\Gamma _{3R}^{(4)}(0,L)=\frac{\lambda }{1+\lambda \Sigma
_{R}(3,L)}. \label{g3L}
\end{equation}
Fig.~1 shows the renormalized single bubble (in units of $m$), $S
= \Sigma _{R}(3,L) / m$, as a function of $L$ (in units of
$m^{-1}$). We see from this plot and from Eq.~(\ref{sigmaR3}) that
$lim_{L\rightarrow \infty}\Sigma _{R}(3,L)=0$ and that $\Sigma
_{R}(3,L)$ diverges as $L\rightarrow 0$. This means that (see
Eq.~(\ref{g3L})) the fixed coupling constant $\lambda$ corresponds
to the large-$N$ renormalized effective coupling in absence of
boundaries and that, independently of the value of $\lambda$, the
system presents ultra-violet asymptotic freedom. Moreover, $\Sigma
_{R}(3,L)$ has a zero at $L = L^{(c)}_{min} \simeq 2.07$, for
which $g(3,L) = \lambda$.

\begin{figure}[h]
\includegraphics[{height=6.0cm,width=8.0cm}]{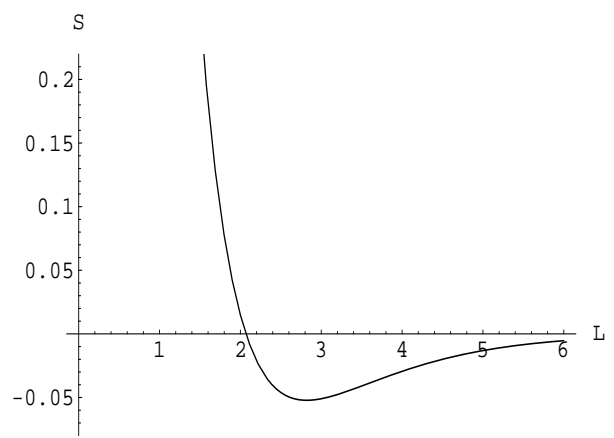}
\caption{Renormalized single bubble (in units of $m$), $S=\Sigma_R
(3,L) / m$, as a function of the length $L$ (in units of
$m^{-1}$).}
\end{figure}

The interesting point is to investigate what happens for {\it
finite} values of $L$, when we start from $L$ near $zero$ (in the
region of asymptotic freedom). In Fig.~2, we have plotted the
relative coupling constant $y=g(3,L)/\lambda$ as a function of $L$
(in units of $m^{-1}$), taking four different values of $\lambda$
(also in units of $m^{-1}$). For the smallest value considered
($\lambda =0.2$), we only see asymptotic freedom with $g(3,L)$
varying from $0$ (for $L=0$) up to $\lambda$ (for $L=\infty$).
Increasing the value of $\lambda$ (e.g. $\lambda =10$ and $\lambda
=15$ in Fig.~2), all the curves present a peak at
$L=L^{(c)}_{max}\simeq 2.82$, the point at which $\lambda /
g(3,L)$ reaches its minimum value. This minimum value of $\lambda
/ g(3,L)$ vanishes for $\lambda = \lambda_c \simeq 19.16$
(corresponding to the full line in Fig.~2) and so the effective
coupling constant $g(3,L^{(c)}_{max})$ diverges. This means that,
for this value of $\lambda$, the system is absolutely confined in
the slab of thickness $L^{(c)}_{max}$. Taking greater values of
$\lambda > \lambda_c$, the minimum of $\lambda / g(3,L)$ becomes
negative, $g(3,L)$ goes to $\infty$ at a value $L^{(c)} <
L^{(c)}_{max}$ and the system would be confined in a thinner slab.
However, as $\lambda$ goes to $\infty$, $L^{(c)}$ tends to a lower
bound equal to $L^{(c)}_{min}$; in other words, for $\lambda_c
\leq \lambda < \infty$ the system is confined in a slab of
thickness $L^{(c)}$ lying in the interval $\left( L^{(c)}_{min},
L^{(c)}_{max} \right]$.

\begin{figure}[h]
\includegraphics[{height=6.0cm,width=8.0cm}]{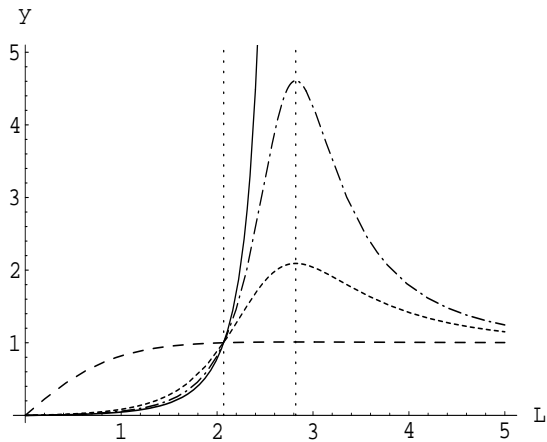}
\caption{Plots of the large-$N$ relative effective coupling
constant, $y=g(3,L)/\lambda$, as a function of $L$ (in units of
$m^{-1}$) for four different values of the fixed coupling constant
$\lambda$ (in units of $m^{-1}$): $\lambda = 0.2$ (dashed line),
$\lambda = 10$ (dotted line), $\lambda = 15$ (dotted-dashed line)
and $\lambda = 19.16$ (full line). The dotted vertical lines,
passing by $L^{(c)}_{min} \simeq 2.07$ and $L^{(c)}_{max}\simeq
2.82$, are plotted as a visual guide.}
\end{figure}

Searching for a physical interpretation of our results, we shall
take the fixed coupling constant $\lambda = \lambda_c$ implying
that the physical confining length is $L=L^{(c)}_{max}$, the
maximum conceived confining length. In order to estimate the value
of $L_0$, we need to fix a value of $m$. We choose $m= 3.25\times
10^{11}cm^{-1}$ ($\sim 6.5 Mev$), corresponding to an average
between typical $u$- and $d$-quark masses (recall that the quark
content of the proton is $\{ uud \}$) \cite{booklet}. This choice
gives a confining length $L^{(p)}_{0}\simeq 8.67\times
10^{-12}cm$. This value should be compared with an estimated
classical proton diameter, $2 R_p \sim 6.92\times 10^{-12}cm$
\cite{note}; notice that this value is about $10^2$ times larger
than the experimentally measured proton charge radius \cite{K}.

Under the assumption that the Gross-Neveu model can be thought
about as a model sharing some of the properties of physically
meaningful theories (in the sense of experimentally testable), we
can say that we have exhibited in a direct way the properties of
ultraviolet asymptotic freedom and confinement, characteristics of
strong interactions. This could be contrasted with the fact that
in $QCD$, although asymptotic freedom is a well established
property on theoretical grounds, confinement has been confirmed
only by computer simulations in lattice field theory. Confinement
has never been, to our knowledge, analytically proved.

It is worth mentioning that, in Ref.~\cite{JMario}, we have
analyzed the dependence of the coupling constant on $L$ for the
large-$N$ $\phi^4$ model, the bosonic counterpart of the
Gross-Neveu model studied here. In that case of bosons, no
confinement is found at all, but only asymptotic freedom is
encountered. Comparison of the results derived in
Ref.~\cite{JMario} with those of the present work suggests that
the effect of confinement in hadronic matter may be mainly due to
the fermionic nature of the quarks.

Finally, we would like to point out that the behavior described in
this work should not be heavily dependent on the confining
geometry. One expects to have ultraviolet asymptotic freedom in
any case and partial or absolute confinement, depending on the
value of $\lambda$. A thorough study on the subject with different
geometries, and including temperature, is in progress and will be
presented elsewhere.

This work was partially supported by CNPq, a Brazilian Agency. One
of us (APCM) is grateful for the kind hospitality of the Instituto
de F\'{\i}sica, Universidade Federal da Bahia, where part of this
work has been done.


\begin{references}

\bibitem{GN}  D. J. Gross and A. Neveu, Phys. Rev. D {\bf 10}, 4410
(1974).

\bibitem{Jarrao}  C. de Calan, P. A. Faria da Veiga, J. Magnen, and
R. S\'eneor, Phys. Rev. Lett. {\bf 66}, 3233 (1991).

\bibitem{Gaw1}  K. Gawedzki and A. Kupianen, Phys. Rev. Lett. {\bf 55}, 363
(1985).

\bibitem{Gaw2}  K. Gawedzki and A. Kupianen, Nucl. Phys. B {\bf 262},
33 (1985).

\bibitem{Weinberg1} S. Weinberg, Phys. Rev. D {\bf 56}, 2303 (1997).

\bibitem{parisi}  G. Parisi, Nucl. Phys. B {\bf 100}, 368 (1975).

\bibitem{anp}  Y. Aref\'va, E. Nissimov, and S. Pacheva, Comm. Math.
Phys. {\bf 71}, 213 (1980).

\bibitem{rwp1}  B. Rosenstein and B. Warr, S. Park, Phys. Rev. Lett.
{\bf 62}, 1433 (1989).

\bibitem{Epstein}  P. Epstein, Math. Ann. {\bf 56}, 615 (1903), {\bf 67},
205 (1907); A. Elizalde and E. Romeo, J. Math. Phys. {\bf 30},
1133 (1989); K. Kirsten, J. Math. Phys. {\bf 35}, 459 (1994).

\bibitem{JMario}  A. P. C. Malbouisson and J. M. C. Malbouisson, J. Phys. A:
Math. Gen. {\bf 35}, 2263 (2002).

\bibitem{Ademir}  A. P. C. Malbouisson, J. M. C. Malbouisson, and A. E.
Santana, Nucl. Phys. B {\bf 631}, 83 (2002).

\bibitem{Chodos}  A. Chodos, R. L. Jaffe, K. Johnson, C. B. Thorn, and V. F.
Weisskopf, Phys. Rev. D {\bf 9}, 3471 (1974).

\bibitem{Lutken}  C. A. L\"utken and F. Ravndal, J. Phys. A: Math. Gen.
{\bf 21}, L793 (1988).

\bibitem{Ramond}  P. Ramond, {\it Field Theory: a modern primer}
(Addison-Wesley, Redwood, 1990), Appendix B.

\bibitem{booklet}  K. Hagivara et al, {\it Review of particle physics},
Phys. Rev. D {\bf 66}, 010001 (2002).

\bibitem{note} This estimate was obtained from the proton-electron mass
ratio and from the tabulated classical electron radius, assuming
equal densities.

\bibitem{K} S. G. Karshenboim, Can. J. Phys. {\bf 77}, 241 (1999).
\end{references}
\end{document}